\documentclass[aps,pre,twocolumn,showpacs,superscriptaddress,groupedaddress]{revtex4}  
\usepackage{graphicx}  
\usepackage{dcolumn}   
\usepackage{bm}        
\usepackage{amssymb}   
\usepackage{amsmath}   
\usepackage{verbatim}
\usepackage{afterpage} 
\usepackage{booktabs}	
\usepackage{floatrow}

\graphicspath{{Graphs/}}

\begin{document}

\title{Dissipation dynamics with two distinct chaotic baths}
\author{Diptanil Roy}
\author{A.V. Anil Kumar}
\affiliation{School of Physical Sciences, National Institute of Science Education and Research, HBNI, Jatni - 752050, India}
\date{\today}

\begin{abstract}
Dissipation using a finite environment coupled to a single harmonic 
oscillator have been studied quite extensively. We extend the study by looking at the dynamics of the dissipation 
when we introduce a second bath of $N$ identical 
quartic systems different from the 1st bath. We look at the energy flow into the environment as a function of the 
chaotic parameters of the bath and also try to develop a 
linear response theory to describe the system. The energy flow is always more to the more chaotic system irrespective of the 
initial energy of the baths.

\end{abstract}

\pacs{05.20.Gg, 05.45.Ac, 05.40.Jc}
\maketitle
\section{introduction}
 The dissipation of energy occurs in diverse systems and it is one of the most fundamental processes in 
 dissipative dynamical systems. In the Langevin description of 
 Brownian motion, the damping force together with the randomly fluctating force are used to model dissipation\cite{Cortes,Risken}. 
 These forces relate the effect of the collisions of the system
 with the particles in the thermal bath and are realted by the dissipation-fluctuation theorem\cite{Risken}.
 The dissipative dynamics of physical systems have been studied extensively\cite{Caldeira,Wilkinson,Berry,Ott,Brown1,Brown2,Marchiori,Xavier}.
 In modelling energy dissipation, these studies used a small system couple to the environment.
 The environment can be of two types : one with an infinite collection of modes\cite{Caldeira} and another 
 with a small number of chaotic modes\cite{Wilkinson,Berry,Marchiori,Xavier}. These investigations have led to several 
 interesting results. For example, Wilkinson\cite{Wilkinson} showed that the rate of exchange of energy between the system
 and the chaotic bath depends on the classical motion of the particles comprising the chaotic bath. The rate of energy 
 exchange increases when the bath shows chaotic motion. Later Marchiori and de Aguiar showed that energy dissipation occurs in the chaotic
 regime even if the bath is composed of small number of particles\cite{Marchiori}. Xavier {\it et al.}\cite{Xavier} also improved upon
 the model in \cite{Marchiori} and showed that the damping rate can be expressed in terms of the mean Lyapunov exponent of
 the chaotic bath. Also they have shown that the dissipation is more effective if there is resonance between the system and bath 
 frequencies.
 
 In this paper, we extend this study to the case of a small system coupled to multiple baths. The purpose of this investigation
 is to study the dissipative dynamics of the system if it is connected to two distinct baths with different chaoticity. We use linear
 response theory following the approach presented in \cite{Marchiori}. We also carry our numerical simulations and compare 
 the theoretical results with those obtained from simulations.
 
\section{The Model}

Our model comprises of a particle, with generalised coordinates (\textit{q, p}) in a one dimensional harmonic oscillator potential with angular frequency $\omega_0$. The 
particle is connected to two baths comprising of $N$ quartic oscillators each. The two baths are not connected to each other, but are coupled indirectly through the harmonic 
oscillator. The quartic oscillators are two dimensional and the $n_{th}$ particle of the $i_{th}$ bath is characterised by the generalised 
coordinates ($\vec{q_{n_i}} = x{_{n_i}}\hat{i} + y{_{n_i}} \hat{j}$, $\vec{p_{n_i}} = p_{x{_{n_i}}}\hat{i} + p_{y{_{n_i}}} \hat{j}$). The interaction between the baths 
and the system is through a coupling term of the form $H_I$, where $\lambda_N = \frac{\lambda}{\sqrt{N}}$ is a measure of the effective coupling \cite{Marchiori}. The Hamiltonian govering 
the dynamics of the system can be written as
\begin{equation}
H = H_{HO} + H_{E_1} + H_{E_2} + H_I \label{Hamiltonian}
\end{equation}
\begin{eqnarray*}
	H_{HO} & = & \frac{p^2}{2m} + \frac{m\omega_0^2q^2}{2}\\
	H_{E_i} & = & \sum_{n=1}^{N}\left[ \frac{p_{{x}_{n_i}}^2 + {p_{y_{n_i}}^2}}{2} + \frac{a_i}{4}(x_{n_i}^4 + y_{n_i}^4) + \frac{x_{n_i}^2y_{n_i}^2}{2}\right] \\
	H_I & = & \sum_{n=1}^{N}\lambda_Nq(x_{n_1} + x_{n_2})
\end{eqnarray*}

The total Hamiltonian is conservative, but the system can exchange energy with the baths and hence behave dissipatively. The non-linear dynamics of the baths is determined 
completely by the parameters $a_i$ where the systems are integrable for $a_i = 1.0$, mildly chaotic for $a_i = 0.1$ and strongly chaotic for $a_i = 0.01$. We numerically 
solve the $8N+2$ equations of motion arising from the Hamiltonians $H_{HO}$, $H_{E_1}$, $H_{E_2}$ and $H_I$ respectively and compare our analytical results to them. The chaotic 
dynamics of the baths acts as a sink of energy for the system. A temperature for the environment can be defined by using the general equipartition theorem, where every quadratic 
degree of freedom has a contribution of $k_BT/2$ to the total energy $E$. It is clear here that the $T$ represents the equilibrium temperature.
\pagebreak

\onecolumngrid

\begin{figure}[h]
	\includegraphics[width = 0.45\textwidth]{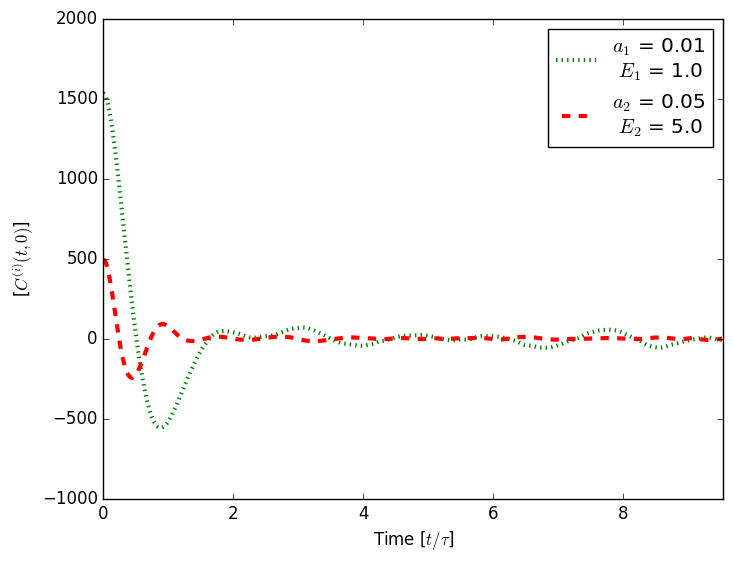}
	\includegraphics[width = 0.45\textwidth]{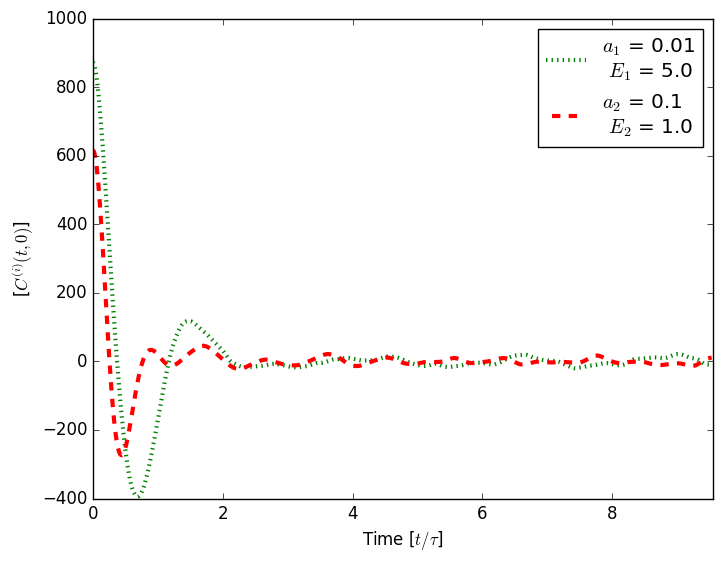}
	\caption{\label{fig:Correlation} The correlation functions $C_N^{(i)}(t,0)$ for the two sets of baths for different set of initial parameters. $C_N^{(i)}$ deviates 
		from the delta function approximation in the mixed and integrable regimes.}
\end{figure}

\twocolumngrid

Before we proceed to the results for our model, it is imperative to discuss the behaviour of a particle in harmonic oscillator potential coupled to only one bath. We will compare 
and contrast our results to those provided in \cite{Marchiori,Xavier}. When $a = 1.0$, the system is in the integrable regime and the environment has very 
little effect, causing a slight decrease in the oscillator energy. With $a=0.5$, the energy loss is more and there are slightly pronounced oscillations. The equilibrium 
energy value for such cases is $0.8 E_0$ \cite{Marchiori}. For example, for a bath with $N = 100$ QOs \cite{Marchiori,Xavier} and $a = 1$, the energy of the system 
decreases a bit from $E_S(0)$ and performs tiny oscillations around the decreased value. For $a=0.1$, the energy decreases further with more prominent oscillations with 
increasing time. For $a = 0.01$, the situation however is completely different. The energy decay for such a system is exponential and can be 
fit to $E_S(t) \approx E_S(0)e^{-\gamma t}$ where $\gamma$ is a function of the bath properties and $\omega_0$.  

For small values of $N$, the indirect coupling does not have much effect on the system unless the coupling is strongly chaotic. For example, when 
the coupling parameter is set in the chaotic regime, we cannot characterise the behaviour by a single realisation for $1 \leq N \leq 6$ because of large 
fluctuations. For $7 \leq N \leq 20$ however, the onset of dissipation and the transition to an exponential decay are apparent, though the energy fluctuation 
is still large. With increasing $N$, the system starts to behave differently, allowing dissipation to occur. The effect of different values of the coupling are discussed above. 

As is apparent from the numerical results, the value of $E_S(t)$ for large N becomes independent of $N$ over large times. This is possible only when the 
coupling term $\lambda_N$ falls with $N$ as $N \rightarrow \infty$. For one bath, it has been shown in \cite{Marchiori} that $\lambda_N = \frac{\lambda}{\sqrt{N}}$ using 
Linear response theory. For large $N$, the system equilibrates with the environment and the equilibrium energy distribution is Boltzmann-like. This allows us to define a 
temperature; however it is not possible in the regular or mixed regimes.

\section{Theory}

We will try to understand the flow of energy to the two different baths using linear response theory (LRT) \cite{Kubo}. We will follow the formalism developed 
in \cite{Marchiori} to write the final equation of motion for the system. 
The \textbf{HO} in hamiltonian \ref{Hamiltonian} satisfies the following equation of motion
\begin{equation}
\ddot{q} + \omega_0^2q = -\frac{\lambda_N}{m}\sum_{n = 1}^{N}\{x_n^{(1)}(t) + x_n^{(2)}(t)\} \equiv -\frac{\lambda_N}{m}X(t) \label{Eqofmotion}
\end{equation}
We consider the \textbf{HO} to be perturbed by an external force given by $-\frac{\lambda_N}{m}X(t)$. Under the assumption that the bath coordinates are chaotic, we can 
replace $X(t)$ with $\langle X(t) \rangle$. LRT has been used to determine $\langle X(t) \rangle$ in terms of the \textit{response function}. The dynamics of the system is 
determined by the Liouville equations.
The integral form of the Liouville equation is given by 
\begin{equation}
\rho(t) = e^{i(t-t_0)L_0}\rho(t_0) + i\int_{t_0}^{t}e^{i(t-s)L_0}L_I(s)\rho(s)ds \label{Liouville}
\end{equation}
where $L_0$ and $L_I$ are the Liouville operators.
The Hamiltonian \ref{Hamiltonian} can be estimated using 
\begin{equation}
H = H_E(Q,P) + H_I(Q, P, t)
\end{equation}
where $Q$ and $P$ represent the entire set of canonical variables of the environment and $H_I(Q, P, t) = A(Q, P)\chi(t)$ where $A(Q,P)$ is the corresponding 
displacement ($\hat{A} = -\frac{\partial}{\partial x}\hat{H}(q,p;x)$).
Under the approximation of a small perturbation, Eq. \ref{Liouville} can be expanded to $1^{st}$ order in $L_I$ as 
\begin{equation}
\rho(t) = \rho(t_0) + \int_{t_0}^{t}e^{i(t-s)L_0}\{H_I(s), \rho(t_0)\}ds \label{LiouvilleApp}
\end{equation}
\pagebreak

\onecolumngrid

\begin{figure}[h]
	\includegraphics[width = 0.45\textwidth]{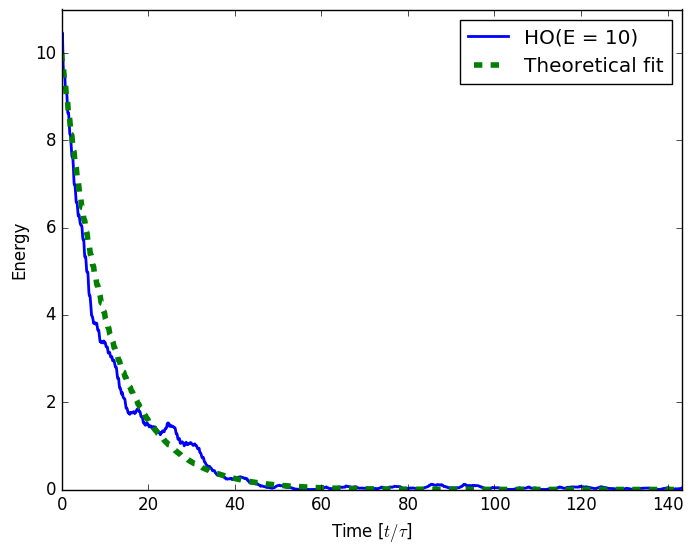}
	\includegraphics[width = 0.45\textwidth]{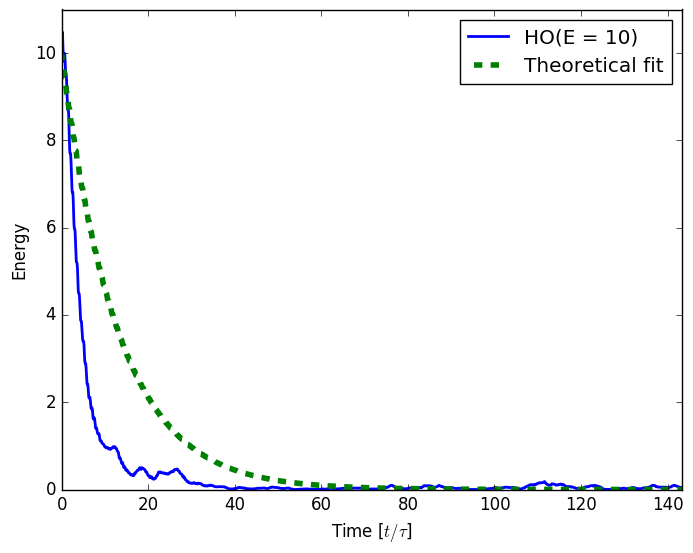}\\
	\includegraphics[width = 0.45\textwidth]{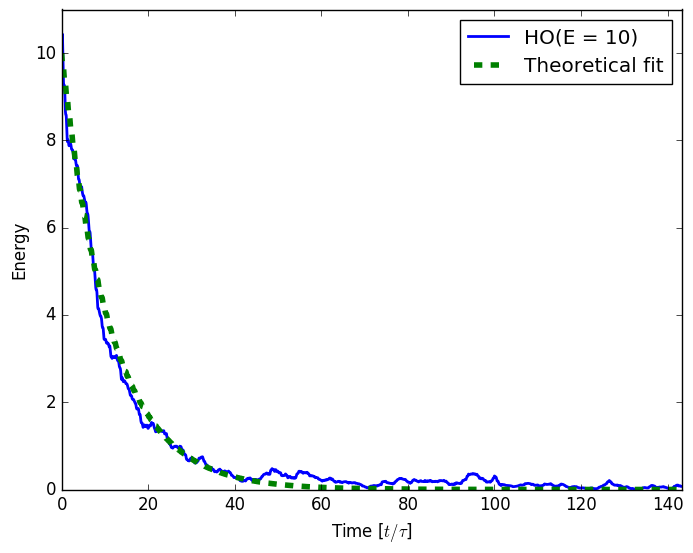}
	\includegraphics[width = 0.45\textwidth]{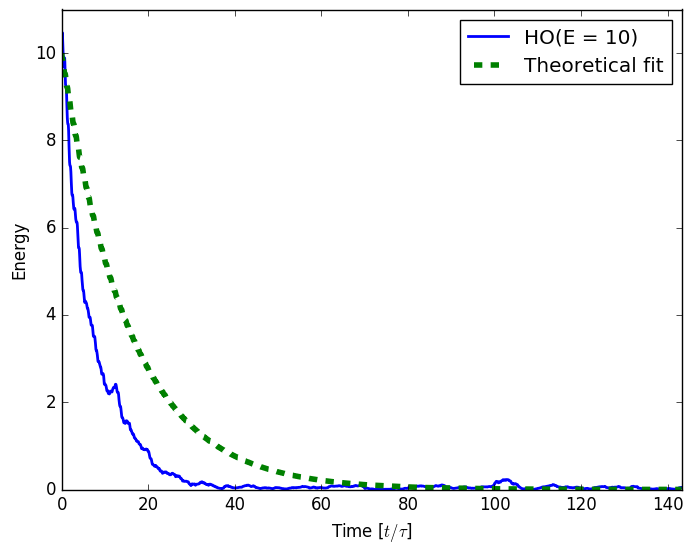}
	\caption{\label{fig:Theoretical} The numerical and theoretical curves for 4 different sets of initial parameters. 
		(a) $a_1 = 0.01$; $a_2 = 0.05$; $E_1 = 1$; $E_2 = 5$ (b) $a_1 = 0.01$; $a_2 = 0.05$; $E_1 = 5$; $E_2 = 1$ 
		(c) $a_1 = 0.01$; $a_2 = 0.1$; $E_1 = 5$; $E_2 = 1$ (d) $a_1 = 0.01$; $a_2 = 0.1$; $E_1 = 1$; $E_2 = 5$}
\end{figure}

\twocolumngrid

The ensemble average $\langle X(t) \rangle$ can thus be written as
\begin{equation}
\langle X(t) \rangle = \left\langle  X(t) \right \rangle_0 - \lambda_N \int_{0}^{t}\phi_{XX}(t-s)q(s)ds
\end{equation}
where  $\left\langle  X(t) \right \rangle_0$ is equal to $0$ due to the parity of $H_E$.
Therefore, Eq. \ref{Eqofmotion} reduces to
\begin{equation}
\ddot{q} + \omega_0^2q \approx \frac{\lambda_N^2 N}{m}\int_{0}^{t}(\phi_{xx}^{(1)} + \phi_{xx}^{(2)})q(s)ds\label{EqofMotion2}
\end{equation}
Here $\phi_{XX}(t-s) = \langle \{X(t), X(s)\} \rangle_0$ and from \cite{Marchiori}, it follows that
\begin{equation}
\phi_{XX}(t-s) = N(\phi^{(1)}_{xx} + \phi_{xx}^{(2)})
\end{equation}
Here, $\phi_{xx}^{(i)}$ is the response function obtained in \cite{Bonanca} and is given by
\begin{multline}
\phi_{xx}^{(i)}(t-s) = \frac{5}{4E_{QS}^{(i)}(0)}\langle x^{(i)}(t)P_x^{(i)}(s)\rangle_0 \\ + \frac{t-s}{4E_{QS}^{(i)}(0)}\langle P_x^{(i)}(t)P_x^{(i)}(s)\rangle_0
\end{multline}
or
\begin{equation}
\phi_{xx}^{(i)}(t-s) = \frac{5}{4}\frac{d}{ds}C_N^{(i)}(t,s) + \frac{t-s}{4}\frac{d^2}{dsdt}C_N^{(i)}(t,s)
\end{equation}
where $C_N^{(i)}(t,s)$ is the correlation function given by
\begin{equation}
C_N^{(i)}(t,s) = \left \langle \sum_{n = 1}^{N}\frac{x_n^{(i)}(t)x_n^{(i)}(s)}{E_{QS}^{(i)}(0)} \right \rangle
\end{equation}
Following \cite{Marchiori}, we make this approximation
\begin{equation}
C_N^{(i)}(t,s) \approx \mu_{\bar{E_i}}^{(i)}N\delta(t-s) \label{Approximation}
\end{equation}
where $A_N^{(i)} =  \mu_{\bar{E_i}}^{(i)}N$ is the maximum amplitude of the correlation function. In the time scale of dissipation, this is a valid approximation when 
the quartic oscillators are in the chaotic regime. As we transition to the mixed and integrable regimes, the approximation fits poorly, however we will consider the 
delta function approximation in the mixed regime too for ease of calculations.

From Eq. \ref{EqofMotion2} and Eq. \ref{Approximation}, we obtain 
\begin{equation}
\ddot{q} + \omega_0^2q + \gamma_T\dot{q} = 0 \label{FinalEqofMotion}
\end{equation}
where 
\begin{equation}
\gamma_T = \frac{7\lambda_N^2(\mu_{\bar{E_1}}^{(1)}+\mu_{\bar{E_2}}^{(2)})N}{4m} \label{GammaT}
\end{equation}

The $\gamma_T$ we obtain theoretically compares well with the numerical results (Fig. \ref{fig:Theoretical}) we obtained.

\section{Numerical Simulations}

We have also simulated the time evolution of our system by solving 8$N$+2 equations of motion obtained from the Hamiltonian in Eq. 1 numerically. The second order 
differential equations were solved using fourth order Runge-Kutta algorithm in a microcanonical ensemble. The time step has been taken to be 0.4 in each case. 
For all the simulations, the value of $\lambda$ is taken as $0.01$. The mass $m$ is taken as $1$ for all the oscillators and the angular frequency of the harmonic 
oscillator $\omega_0$ is taken to be $0.3$. The time is measured in terms of the harmonic oscillator time period $\tau \approx 20.93$.

\section{Results and Discussion}

The results for our model are obtained both from Linear response theory and numerical simulations. Figure 2 shows the energy of the 
harmonic oscillator for four different set of initial conditions obtained from both simulations and linear response theory. The parameters
for the harmonic oscillators are same for all the four graphs. The thoeretical and numerical results are in good agreement with each other. The deviations
can be attributed to the $\delta$-fucntion approximation for the correlation function used in the liner response theory. In the case of one chaotic bath, this has 
been improved upon by Xavier {\it et al.}\cite{Xavier} by incorporating the history and frequency depence of the system. We intend to incorporate these in our model and 
the results will be reported elsewhere. Figure 2 confirms that the harmonic oscillator is dissipative and energy flow occuring from the system to the bath. Also we have done 
simulations in which both the baths are not chaotic. In this case, the harmonic oscillator does not dissipate energy considerably into the baths and the energy dissipation happens
only if at least one of the baths is in the chaotic regime.

Next, we have analysed the energy flow from the harmonic oscillator to the chaotic bath. For this, we have calculated the total energy of the system and each bath
at regular intervals. This has been shown in Figure 3 and Figure 4. In Figure 3, we have plotted the total energy of the system and the two baths, both of which are in 
chaotic regime, but with different chaotic parameter and initial energy, versus time. In Figure 3(a) the the more chaotic system has less initial energy and less chaotic system 
has more initial energy. As evident from the figure, more of the dissipated energy from the system goes to the more chaotic system. Figure 2(b) shows an interesting behaviour
in the energy flow. In figure 3, the more chaotic system has more energy and less chaotic system has less energy. So natuarally one would expect more energy to flow to the 
bath which has lesser energy( and less chaotic). However, more of the dissipated energy has flown into the more chaotic system even though its energy is higher than the other bath.

\begin{figure}
	\includegraphics[width = 0.9\textwidth]{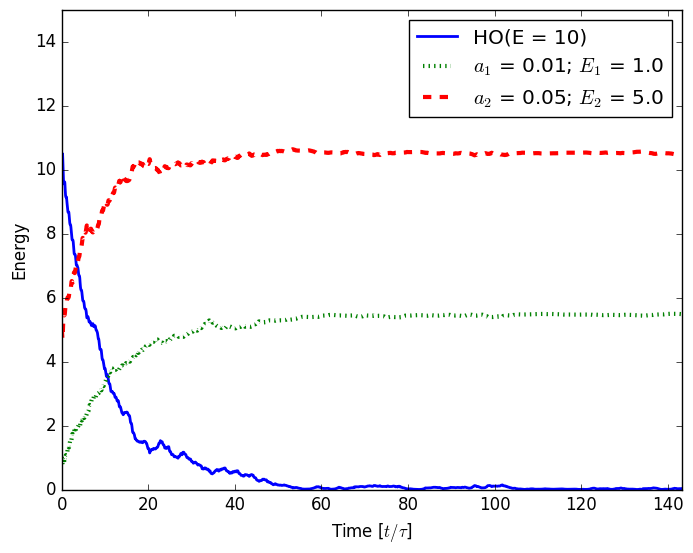}\\
	\includegraphics[width = 0.9\textwidth]{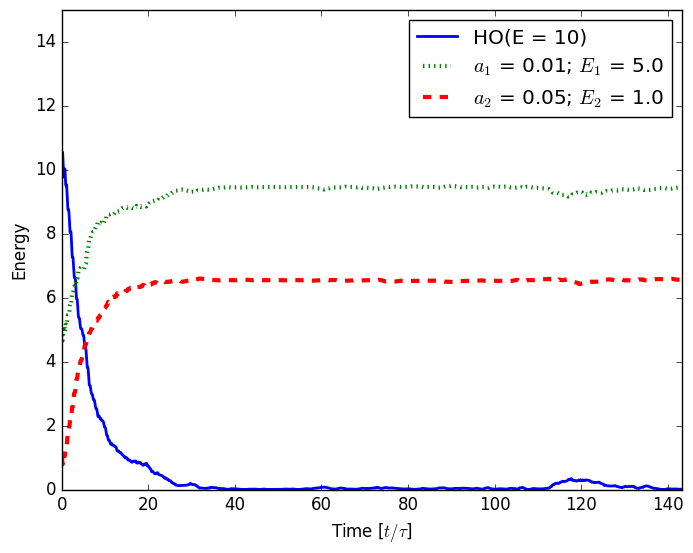}
	\caption{\label{fig:EnergytoBath1} The energy flow into two different sets of baths, both in chaotic regime . The central \textbf{HO} has an initial energy of $10$.}
\end{figure}

\begin{figure}
	\includegraphics[width = 0.9\textwidth]{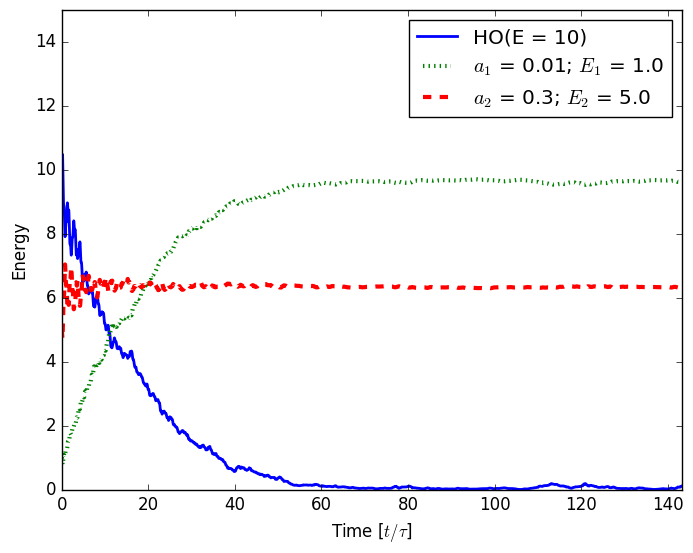}\\
	\includegraphics[width = 0.9\textwidth]{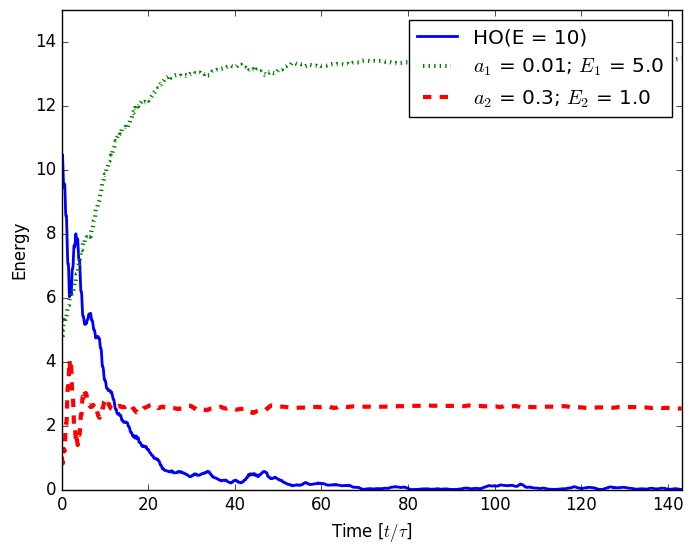}
	\caption{\label{fig:EnergytoBath2} The energy flow into two different sets of baths, one in chaotic and one in mixed regime. The central \textbf{HO} has an initial energy of $10$.}
\end{figure}

Figure \ref{fig:EnergytoBath2} shows two graphs which depicts the enrgy flow into the two different baths, one bath in the chaotic regime and the other in the mixed regime. Here also it
is evident that irrespective of the initial energy of the baths, the energy dissipated from the system flows more into the chaotic system. This has been the case with 
all the simulations, we have done in our model for different parameters. Also, it has been observed that the bath in the mixed regime attained the steady
state much faster than the chaotic bath.

This result is rather surprising and nonintuitive. Using the general equipartition theorem, we can define a temperature for the environment. Therefore, each bath 
contributes $E^{(i)} = N\frac{k_BT}{2}$ to the total energy. Therefore, different initial energies for the two baths essentially means coupling the 
central \textbf{HO} to two baths at two different temperatures, the bath with higher energy being at a higher temperature. In such case, normally one would expect 
more energy to flow to the bath at lower temperarute. This has been the case if the low temperature bath is more chaotic. However, the reverse happens if the low temperature 
bath is less chaotic.

The seemingly counterintuitive result may be explained in terms of the entropy of the system. The more chaotic system will 
have more entropy and the energy will flow 
into the more chaotic bath such that the total entropy of the system will increase. This also causes the more chaotic
system to take more time to reach the steady state, which is evident in Figure 4. More analysis are being 
carried out to quantitatively justify this observed result.

\section{conclusions}

 The dissipative dynamics of a one dimentional harmonic oscillator coupled to two distinct baths has been investigated using 
 linear response theory and numerical simulations. It has been shown that the harmonic oscillator dissipates energy 
 into the baths if the baths are chaotic. The numerical and theoretical results compares well. We have also analysed the 
 dissipative energy flow into the baths from the system. It has been observed that more energy flows into the 
 more chaotic system irrespective of the initial energy of the baths. This rather surprising result may be explained based on 
 the maximization of entropy of the system. One might be able to justify this more precisely by calculating the dynamical 
 entropy of the system for different sets of parameters. 

\end{document}